\documentclass[a4paper,twoside]{article}

\usepackage{epsfig}
\usepackage{subfigure}
\usepackage{calc}
\usepackage{amssymb}
\usepackage{amstext}
\usepackage{amsmath}
\usepackage{amsthm}
\usepackage{multicol}
\usepackage{pslatex}
\usepackage{apalike}

\graphicspath{{./figs/}}

\usepackage{array}
\newcolumntype{P}[1]{>{\centering\arraybackslash}p{#1}}

\usepackage{url}
\usepackage{dblfloatfix}
\usepackage{SCITEPRESS}     

\makeindex

\begin{document}
\title{Building a Cybersecurity Risk Metamodel\\
for Improved Method and Tool Integration}

\author{Christophe Ponsard
\affiliation{CETIC Research Centre, Charleroi, Belgium}
\email{christophe.ponsard@cetic.be}}


\abstract{Nowadays, companies are highly exposed to cyber security threats. In many industrial domains, protective measures are being deployed and actively supported by standards. However the global process remains largely dependent on document driven approach or partial modelling which impacts both the efficiency and effectiveness of the cybersecurity process from the risk analysis step. In this paper, we report on our experience in applying a model-driven approach on the initial risk analysis step in connection with a later security testing. Our work rely on a common metamodel which is used to map, synchronise and ensure information traceability across different tools. We validate our approach using different scenarios relying domain modelling, system modelling, risk assessment and security testing tools.}

\onecolumn \maketitle \normalsize \vfill

\section{\uppercase{Introduction}}

Many domains have become dependent on information or industrial control systems (ICS) for their daily operation. While digital technologies enable systems to be more responsive, automated and competitive, their growing amount, complexity and interconnection increase their attack surface \cite{Trustwave20}. It is becoming an essential concern in many industrial sectors which need to make sure their organisation, services and products are able to cope with this evolution. This has triggered the evolution and creation of many standards and frameworks \cite{NIST-CSF}\cite{CIS}. Europe is also aware of this and has taken action with the NIS (Network Information System) directive in an increasing number of essential sections  (transport, energy, water treatment, health, etc.) \cite{NIS-TEXT}.

Deploying security measures requires actions along several lines of defence (protection, reaction, recovery) and the adoption of a strong risk management approach to first identify the risks, evaluate them before deciding on the measures to reduce them to an acceptable level \cite{ISO-risk}. Such measures need to be implemented and tested for their effectiveness either at development time (functional security testing, at deployment time (penetration testing) and at run-time using security monitoring, based on a DevSecOps lifecycle \cite{DevSecOps}.

Risk management is well defined in many domains and a mandatory step in all standards with a degree of detail and adoption varying according to the considered domain. For example, while the information technology (IT) sector benefits from the ISO27K standard series since 2005, while other sectors have only more recent adoption: industrial systems and aeronautics around 2010 and automotive only in 2021. This diversification is positive as it is pushing to cope with the specific aspects of each domain in terms of assets and technologies, such as IT (Information Technology) and/or OT (Operation Technology). However, the diversity also results in more barriers to their understanding and control. At the conceptual level, some risk-related concepts used may be used with slightly different semantics, level of granularity or be domain specific. Considering tool support, it is also more difficult to easily establish and automate mappings between various tools taking part in a DevSecOps toolchain, especially when considering a model-based approach. In the scope of this paper, we will more specifically focus on the output of modelling, risk management and testing tools.

This paper is organised as follow. First Section 2 gives an overview of existing metamodels. Section 3 outlines the proposed conceptual model covering notions such as assets, risks, threats and measures. Then, section 4 presents our validation on existing tools and discusses it in Section 5. Finally, Section 6 concludes and presents our future work.

\section{\uppercase{Survey of Risk Metamodels}}

This section quickly depicts a number of risks metamodels published in the literature in order to identify they commonalities and interesting features to capture in order to build a rich conceptual model able to cope with the needs of various domains, lifecycle activities and the related tooling. We start from more general and simpler metamodels and comment their progressive enrichment.

\subsection{Generic Risk Model}

\begin{figure}[!h]
\centering
\includegraphics[width=\columnwidth]{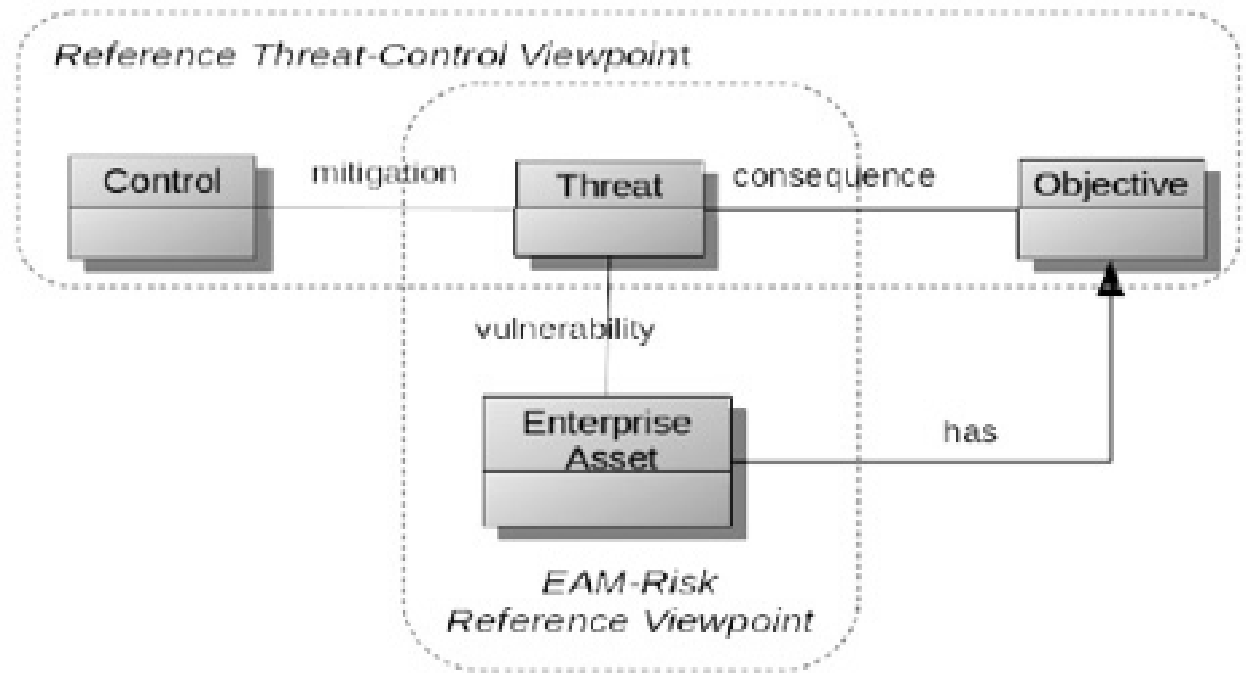}
\includegraphics[width=\columnwidth]{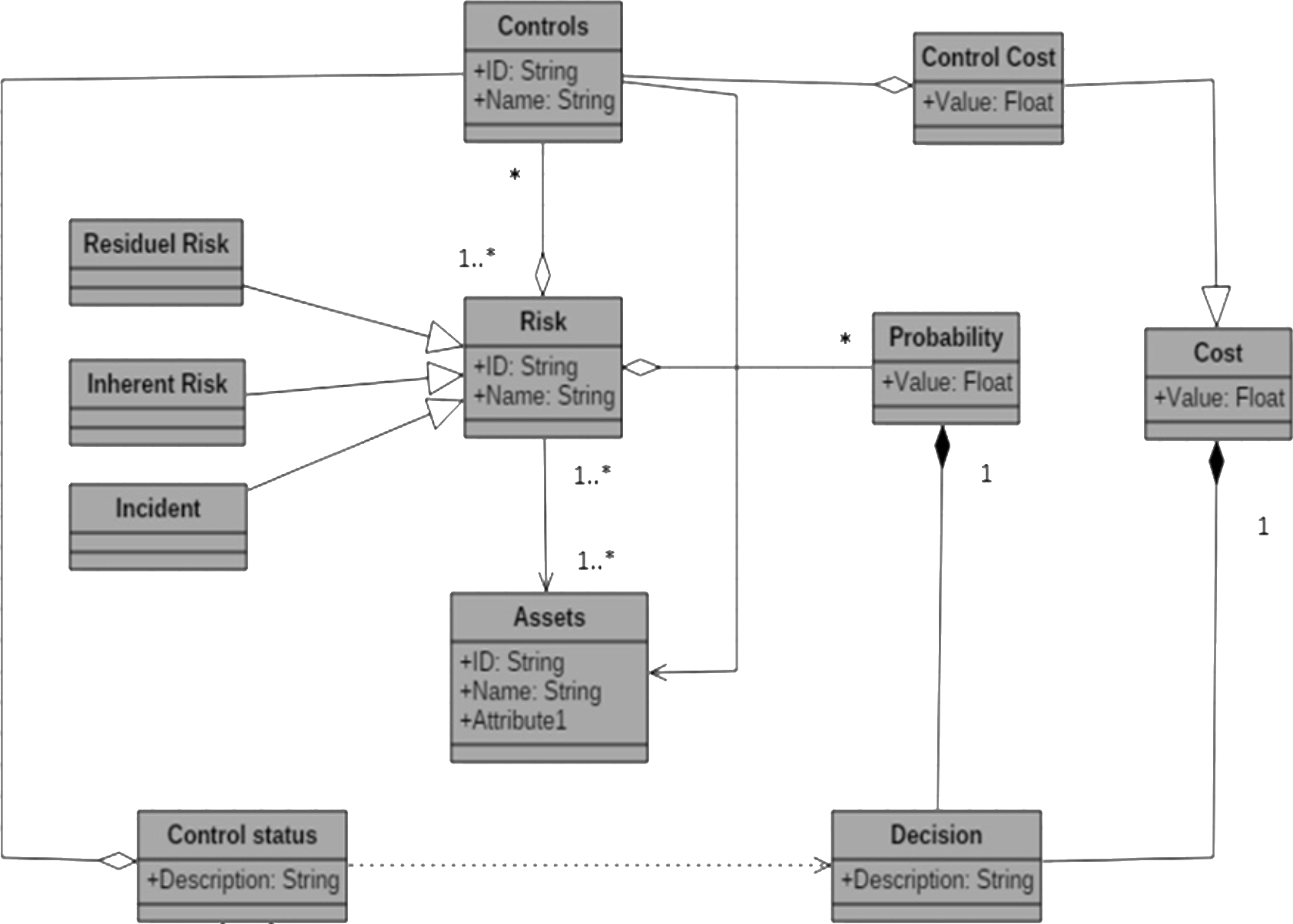}
\caption{Cost-Risk Model from \cite{Chouba18}}
\label{fig:MM-costs}
\end{figure}

Figure \ref{fig:MM-costs} presents a Cost-Risk metamodel developed for optimisation purposes in the context of regulation but not specific to cybersecurity. It reveals interesting core concepts such as asset, objective, threat and control in the top part. In the bottom part, risk (with different flavours such as inherent and residual) is bound to assets and controls.

\subsection{Simple ISO27K Cybersecurity Risk Model}

\begin{figure}[!h]
\centering
\includegraphics[width=\columnwidth]{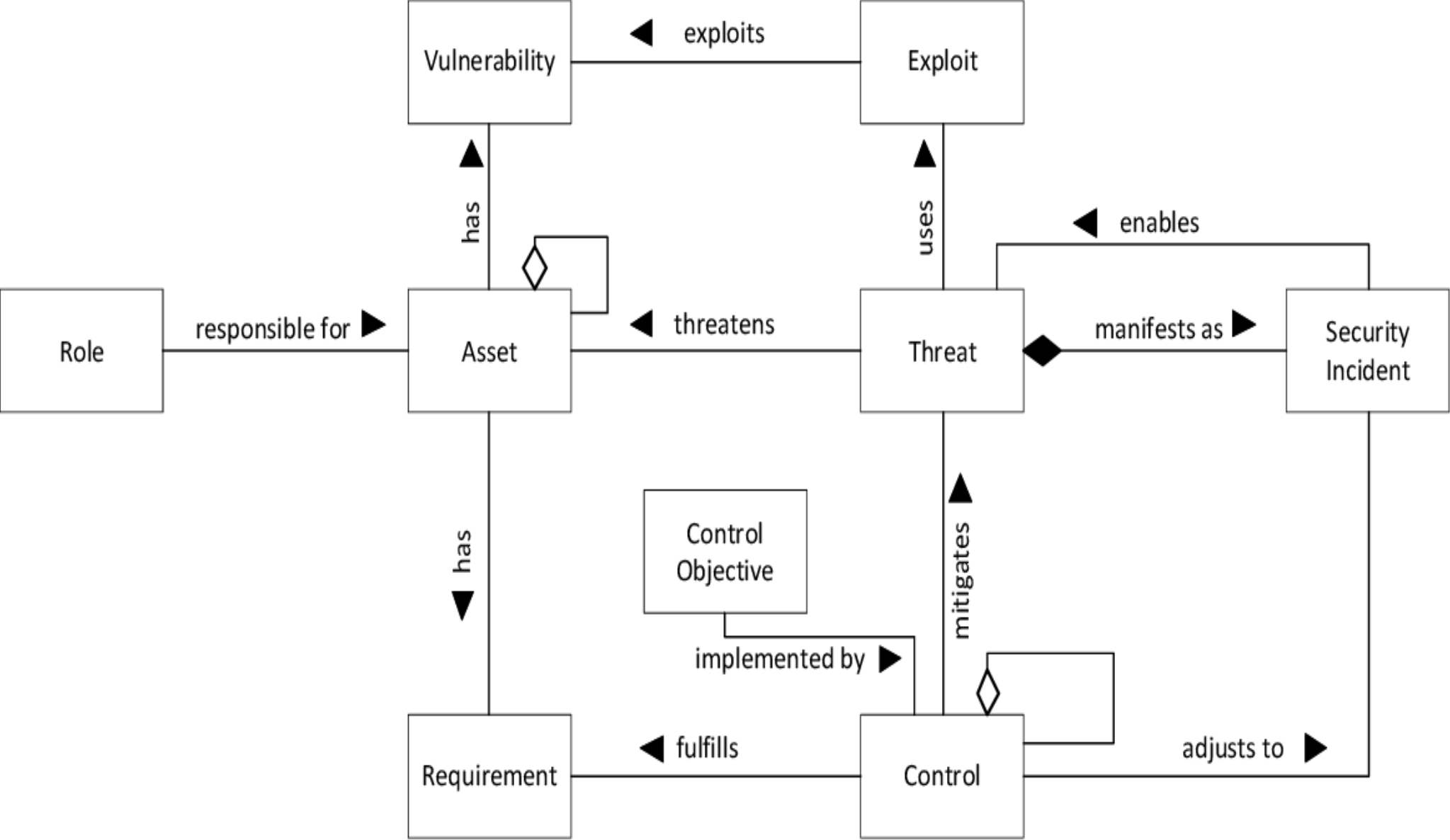}
\caption{Simple Risk Model for ISO27K from \cite{Milicevic10}}
\label{fig:MM-iso27k}
\end{figure}

Figure \ref{fig:MM-iso27k} presents a simple cybersecurity risk model for the ISO 27001 standard \cite{ISO27K} which specifies the requirements on the information security management system based on a risk-management approach \cite{Milicevic10}. Compared to the previous metamodel, specific security concepts such as vulnerability and exploits are added. Requirement is used but is connected with the notion of objective. A responsibility role is also identified. Note that the risk concept is not explicitly modelled here.

\subsection{EBIOS Risk Model}

\begin{figure}[!h]
\centering
\includegraphics[width=\columnwidth]{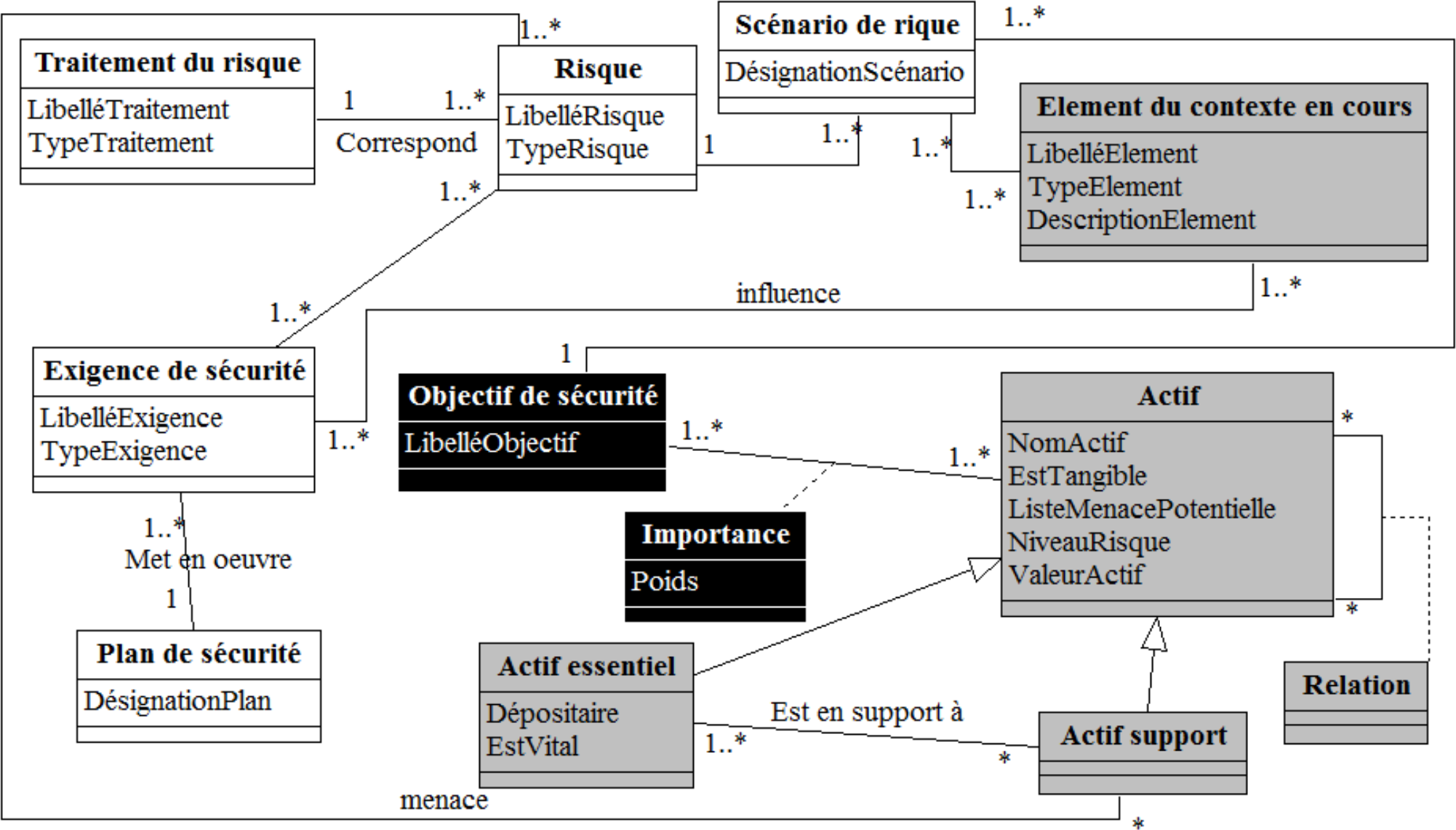}
\caption{EBIOS Risk Model \cite{Akoka18}}
\label{fig:MM-akoka}
\end{figure}

Figure \ref{fig:MM-akoka} depicts a metamodel for EBIOS ("Expression des Besoins et Identification des Objectifs de Sécurité") which is a French method supported by ANSSI, the national cyber security authority of France. It is compliant with the ISO27005 \cite{EBIOS}. The concepts in French cover our assets ("actifs") in two flavours: business ("essentiel") and support. Risk is modelled explicitly and bound with a notion of scenario which can impact a security objective with a quantified impact ("importance"). Threat is capture as a relation linking the risk to the targeted support asset.

\subsection{IEC/ISA 62443 Conduits}

\begin{figure}[!h]
\centering
\includegraphics[width=\columnwidth]{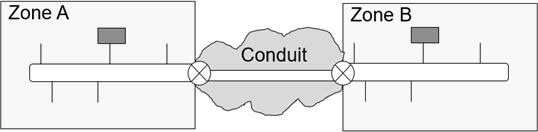}
\caption{Conceptual modelling of zones and conduits in ISO 62443}
\label{fig:ZC}
\end{figure}

Another interesting cybersecurity standard is the IEC/ISA 62443 aiming at securing components or systems used in industrial automation and control systems (IACS) \cite{IEC62443}. Compared to ISO27K is focusing on information technology at the enterprise and information management layers, this standard is also considering operation technology such as control/ supervision and industrial sensors/actuators. An interesting point is the support of partition into zones and conduit depicted in Figure \ref{fig:ZC}. This provides isolation to better organise lines of defence.

\subsection{Risk Model for Security Engineering}

\begin{figure}[!h]
\centering
\includegraphics[width=\columnwidth]{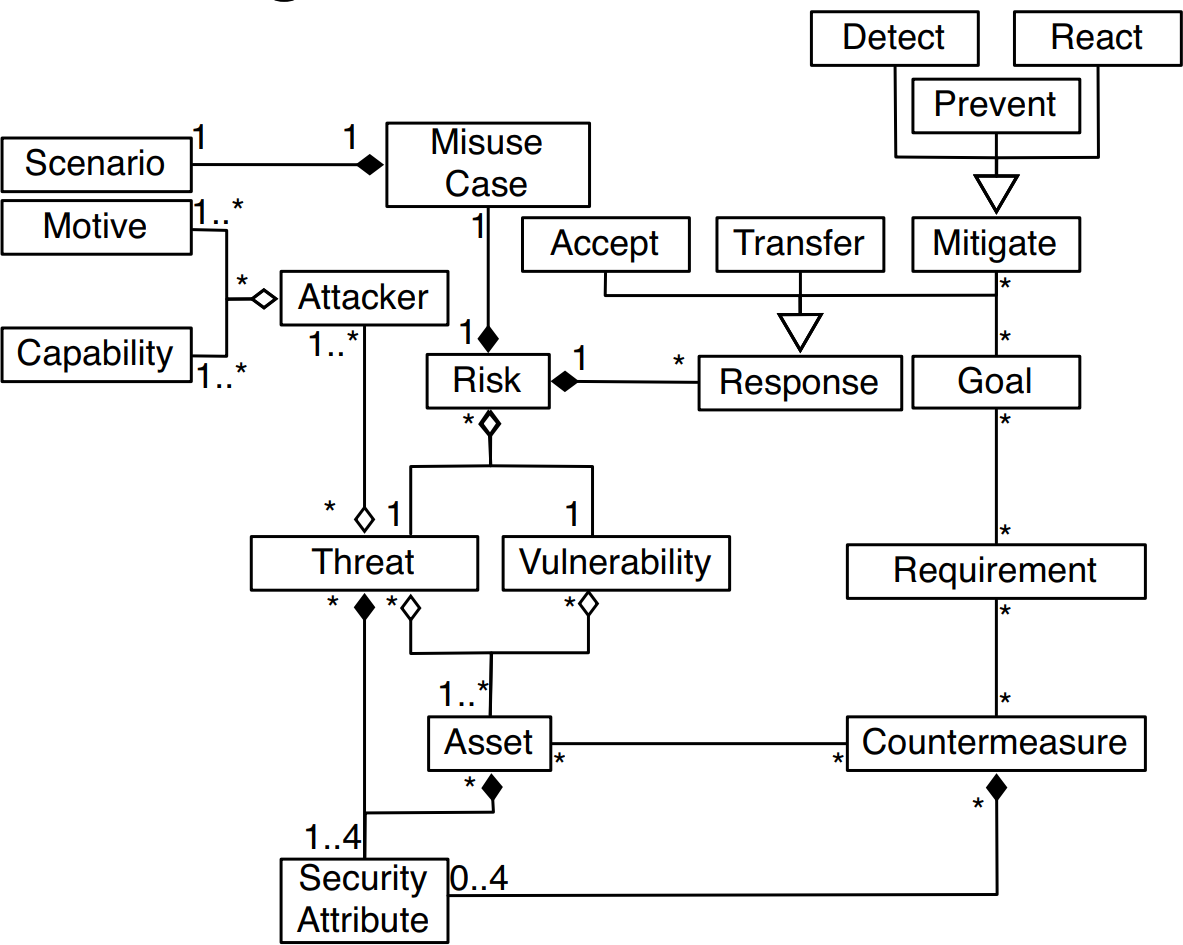}
\caption{Risk metamodel for security engineering from \cite{Faily19}}
\label{fig:MM-faily}
\end{figure}

Figure \ref{fig:MM-faily} presents the risk metamodel from a wider metamodel aiming at proposing a usable security engineering approach and composed of various submodels viewpoints for tasks, goals, risk and responsibilities. It also shares some commonalities with metamodels of requirements engineering tools such as KAOS and i* which are too general to explore here but interesting to use as validation case (see section 4). Additional concepts here relate to the more explicit modelling of the attacker which his capability and motive. Attack scenarios are also structured as more general misuse cases. Measures are named responses and classical strategies such as accept, transfer or mitigate are modelled as specialisations. Countermeasures are more specifically linked to the mitigate strategy.

\subsection{Risk Model for Safety-Security Co-Engineering}

\begin{figure}[!h]
\centering
\includegraphics[width=\columnwidth]{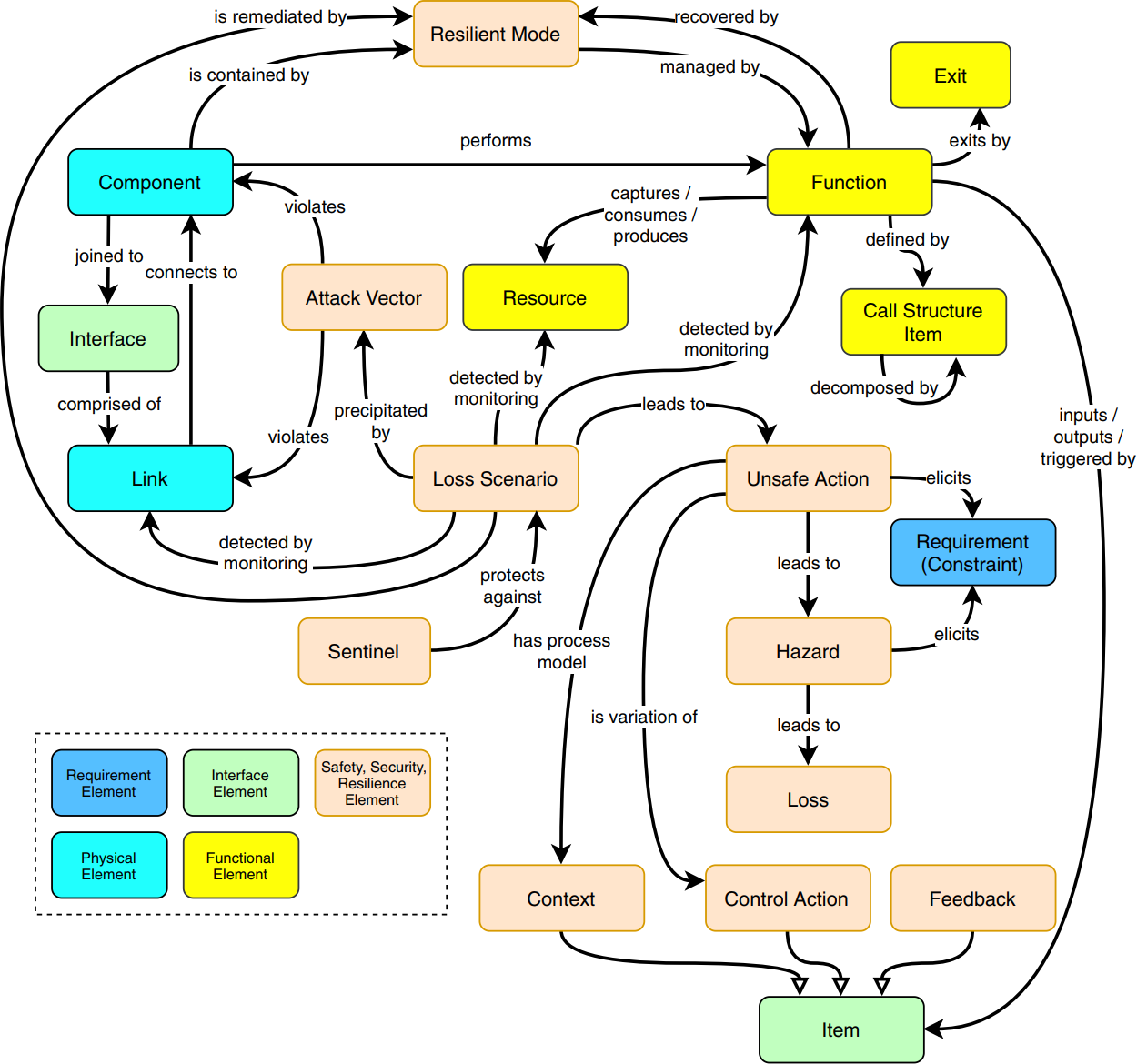}
\caption{Risk metamodel for safety-security co-engineering from \cite{Bakirtzis22}}
\label{fig:MM-baki}
\end{figure}

Finally, we also explore a wider and more recent approach mixing safety and security which share many similarities although the context and also the culture of their respective engineers are quite different. Figure \ref{fig:MM-baki} shows such an ontology build for that purpose \cite{Bakirtzis22}. Compared to our previous metamodel, the concepts are harder to map: assets are captured rather through functions (primary) or component (support). Attack vector relate to a threat and are activated by a loss scenario. The interesting point is the connection with safety through the triggering of unsafe action. As this model is less aligned and not expressed at the same granularity level, it points the limit of trying to encompass multiple target domains.

\section{\uppercase{Consolidated Conceptual Risk Model}}

This section describes our consolidated conceptual risk model build based on the survey performed in the previous section.

\begin{figure*}[!htb]
\centering
\includegraphics[width=0.8\textwidth]{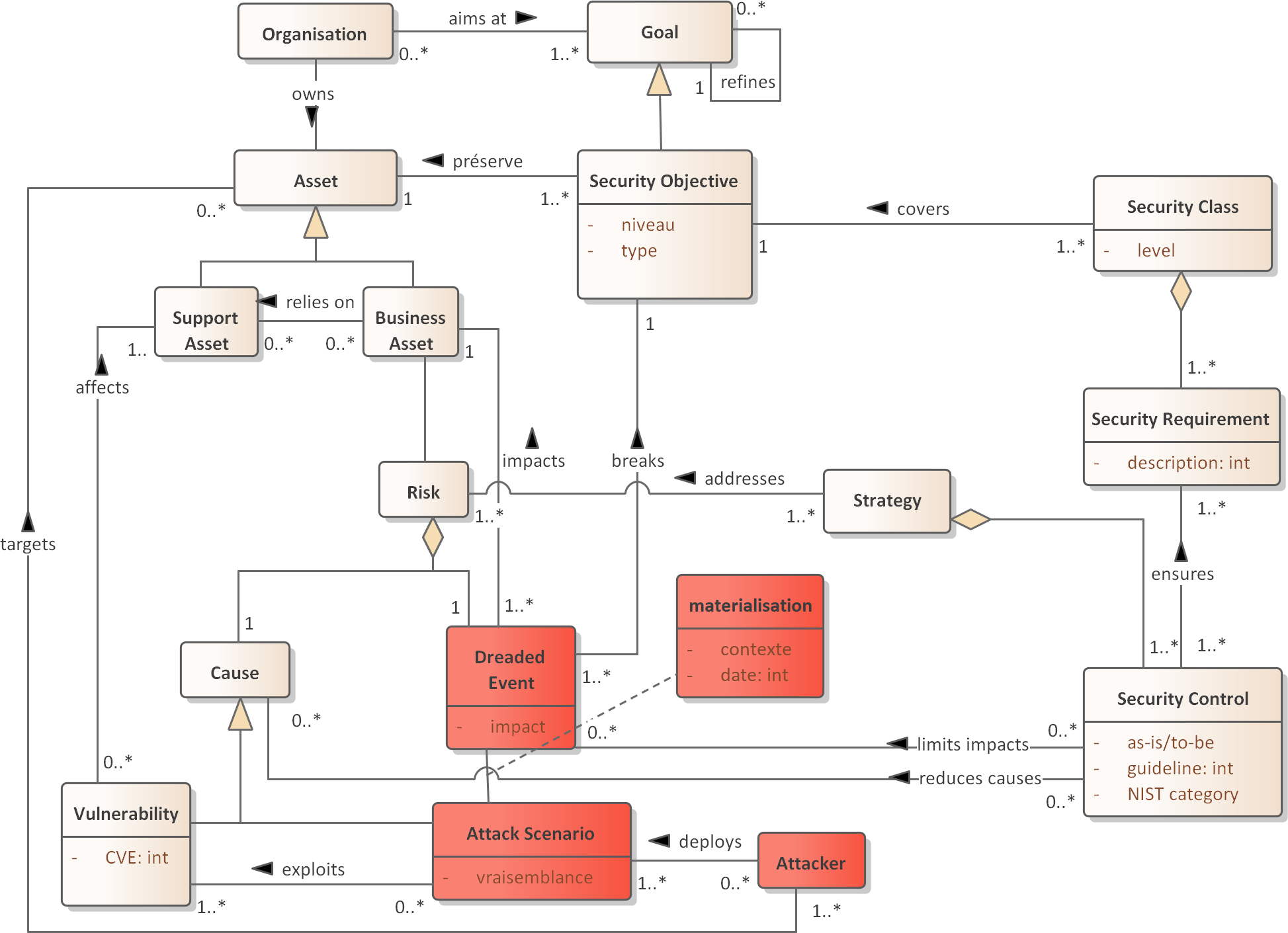}
\caption{Risk Analysis concepts and their relationships}
\label{fig:MM-risks}
\end{figure*}

\begin{figure}[!h]
\centering
\includegraphics[width=\columnwidth]{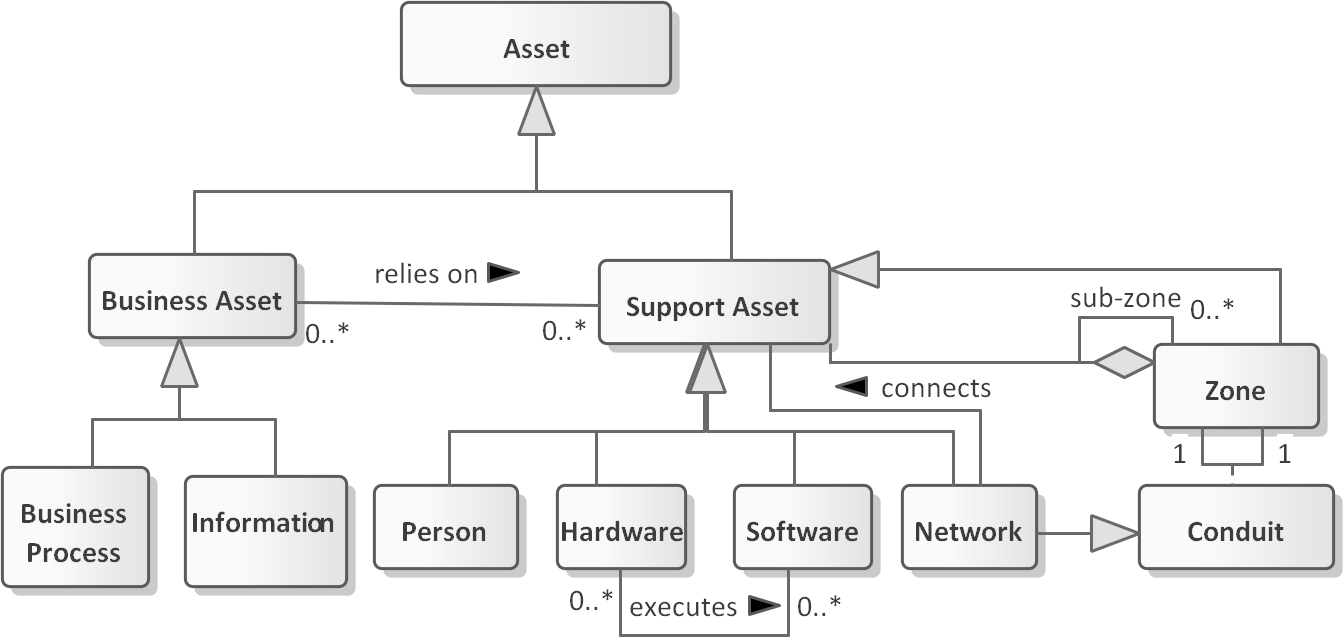}
\caption{Classification of essential and support assets}
\label{fig:MM-assets}
\end{figure}

Figure \ref{fig:MM-assets} structures the business and support assets according to the classification proposed by EBIOS as well as the structuring in zones and conduits of ISO 62443. Figure \ref{fig:MM-risks} describes the articulation between the business assets, the security properties to be guaranteed, the risks and the control measures. The key features are:
\begin{itemize}
\item the part relating to the organisation allows the structured modelling of objectives via goal trees. This makes it possible to link security properties to other system goals and to use goal-oriented requirements engineering notations and methods such as KAOS \cite{vlam09} or i* \cite{Yu97}.
\item In a dual way, the notion of attacker is made explicit and his motivations are captured by anti-goals called here dreaded events (EBIOS term) which are materialised via attack scenarios. This part linked to the modelling of the attacker is represented in a red colour in the figure \ref{fig:MM-risks}.
\item the risk itself is captured according to its dimensions of impact (in connection with the business assets and properties) and feasibility (in connection with the supporting assets which are those that can contain vulnerabilities exploitable by attack scenarios).
\item Finally, security risks are addressed by strategies that implement controls. These are located on different lines of defence specified using the NIST CSF framework \cite{NIST-CSF}.
\end{itemize}

\section{\uppercase{Validation}}

\begin{figure*}[!t]
\centering
\includegraphics[width=0.75\textwidth]{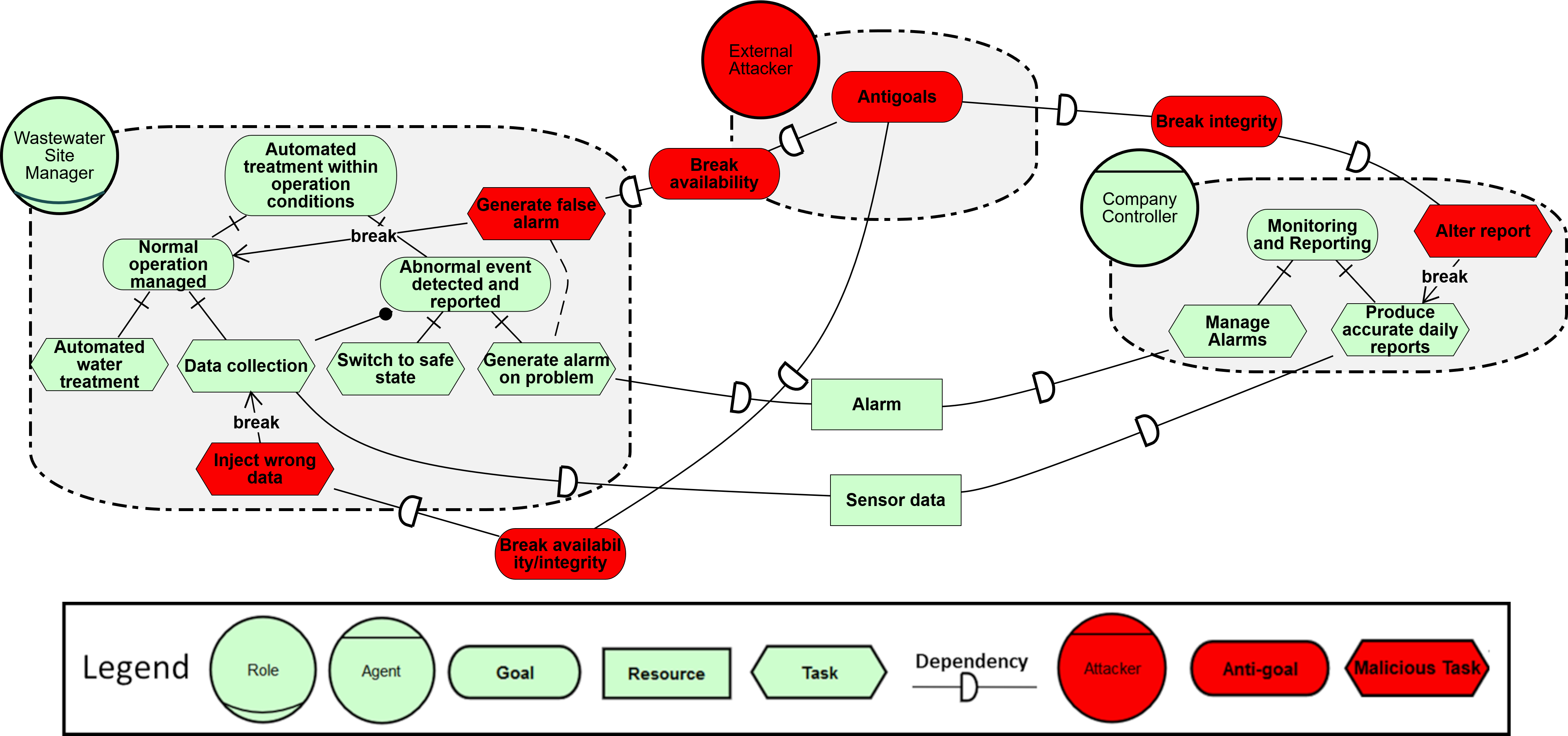}
\caption{piStar modelling of a water treatment facility}
\label{fig:pistar}
\end{figure*}

\begin{figure*}[!b]
\centering
\includegraphics[width=0.75\textwidth]{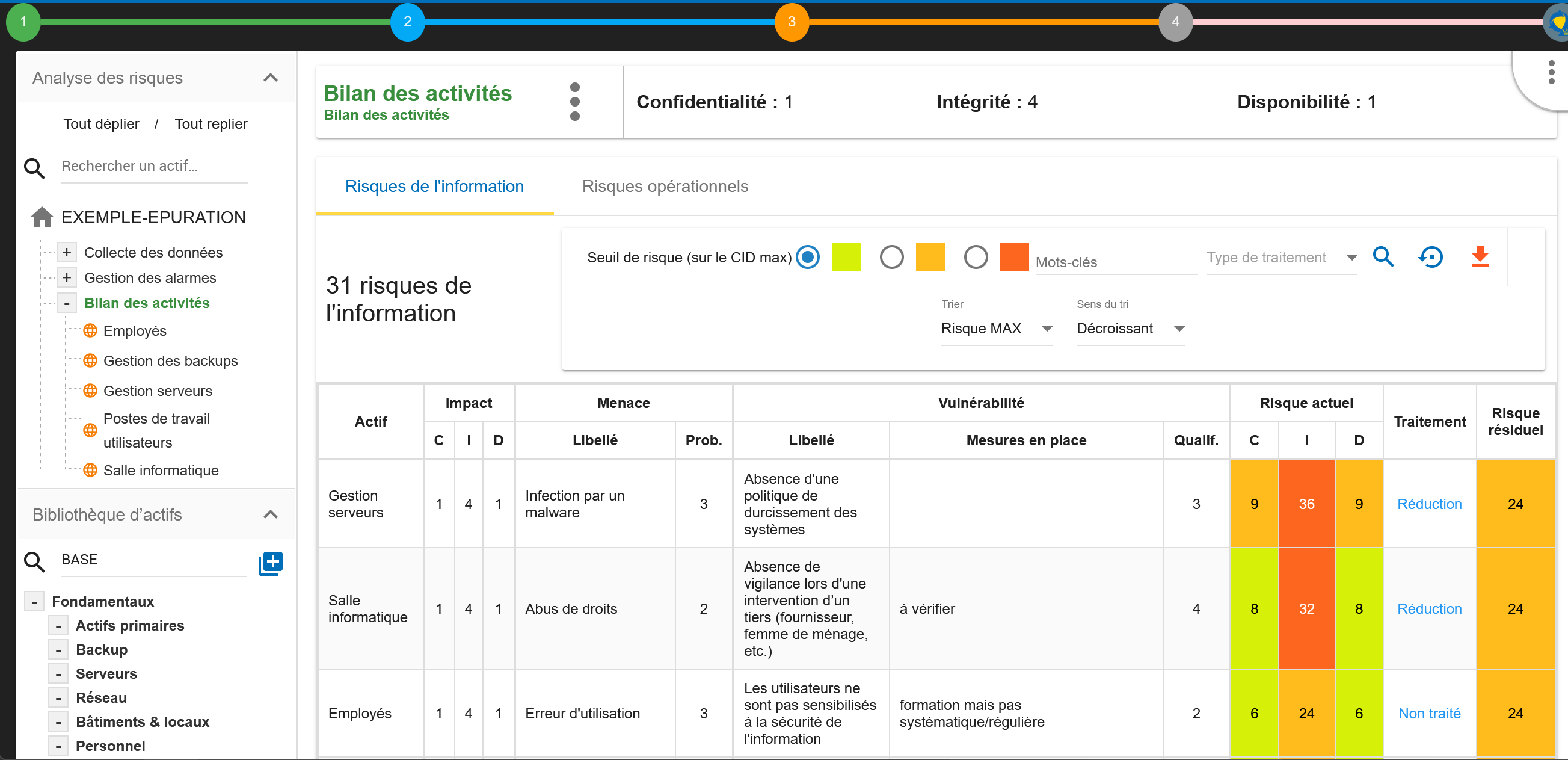}
\caption{MONARC risk assessment tool}
\label{fig:monarc}
\end{figure*}

In order to validate our model, we used a practical approach trying to model or capture different concepts in various tools at specific stage of the DevSecOps process, more specifically:
\begin{itemize}
\item the piStar strategic modelling tool \cite{pistar} already investigated in a previous work [reference removed]
\item the MONARC web-based risk assessment tool \cite{Monarc}
\item the Capella extension for risk assessment \cite{Naouar21}
\item the CYRUS test framework \cite{CYRUS}
\end{itemize}

The method used was either
\begin{itemize}
\item to try to use the tooling and implement some exchange of data, e.g. between piStar and MONARC based on the concept revealed by their diagrams, tables or data format. We used a simple water treatment facility as example.
\item to study an available published metamodel or database schema, in the line of what was done in section 2, e.g. for Capella and CYRUS
\end{itemize}

We then confront the retrieved concepts with our metamodel to identify the mapping and possible limitations or refinement needs.

\subsection{piStar Strategic Tool}

Figure \ref{fig:pistar} depicts the strategic rationale diagram used to capture the business level of a water treatment validation case. The legend shows the full range of concepts used. We can easily map goals, resources (support assets), attackers, anti-goals (dreaded events) and malicious tasks (attack scenarios). The organisation is captured through a collection of agents but the notion of role is not totally captured (although identified earlier) because the person concept is distinct from its role. Those concept are also easy to export using a JSON format proposed in the tool interface \cite{pistar}.

\subsection{MONARC Risk Assessment Tool}

\begin{figure*}[!t]
\centering
\includegraphics[width=0.80\textwidth]{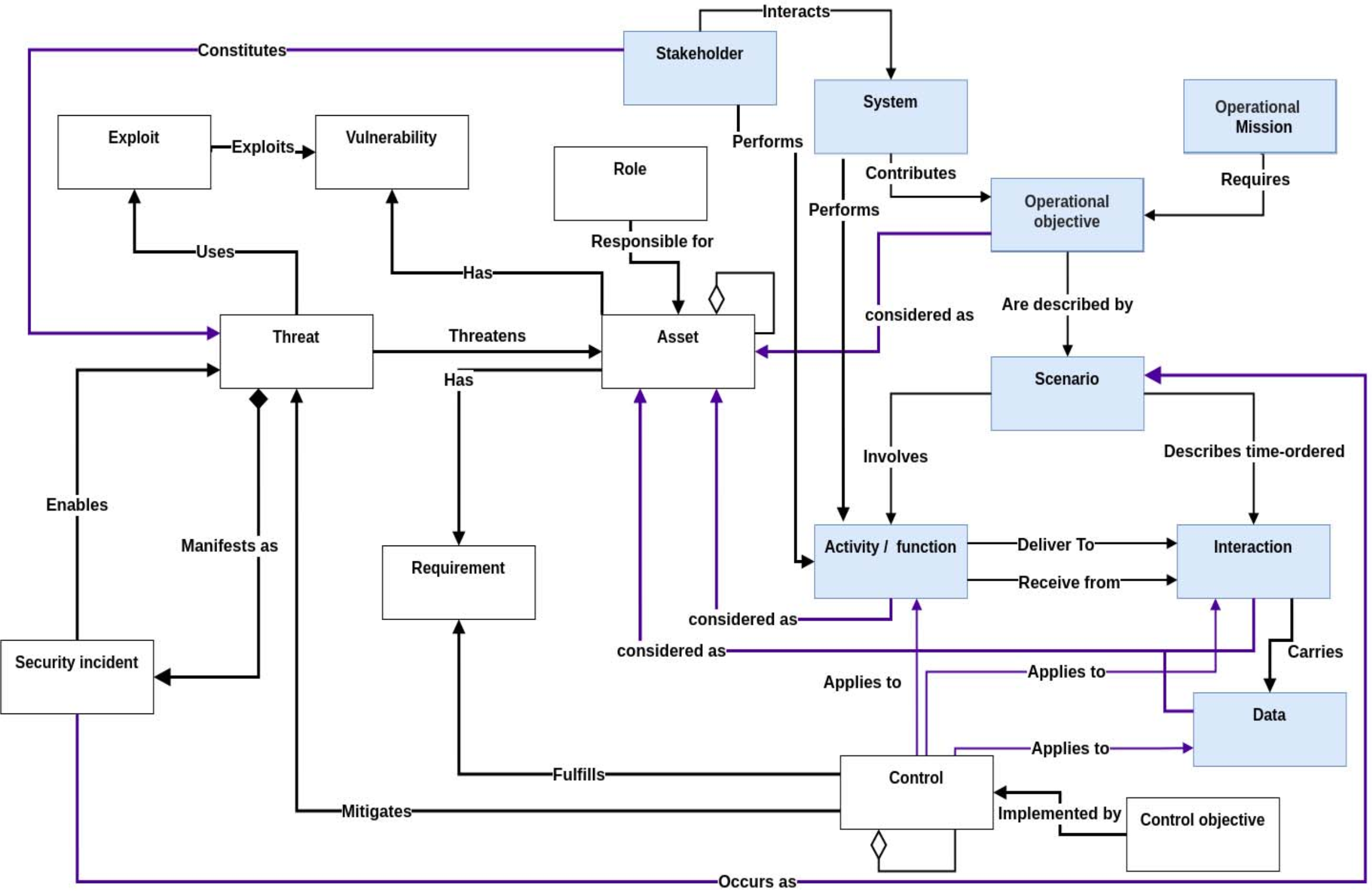}
\caption{Capella metamodel for Risk Assessment}
\label{fig:MM-naouar}
\end{figure*}

MONARC is a method and tool supporting the ISO 27005. Figure \ref{fig:monarc} represents the user interface of the tool depicting the different assets (on the left part. It also feature a more generic level of reusable assets which can be reused. The table shows a list of risks characterised with the related thread, vulnerability and a set of measures than are used to address them with different strategies described earlier (accept, mitigate, transfer,...). The tool also provides a rest interface allowing to create/read/update/delete different concepts through projects endpoints for assets, risks, vulnerabilities, threats which directly match our metamodel. Note however that the relationship between primary assets and support assets is captured through an aggregation link: the primary asset is supposed to be at the top level and support assets at lower levels of the navigation tree. The same structure is reflected at the API through nested assets in the JSON format.

\begin{figure*}[!t]
\centering
\includegraphics[width=0.85\textwidth]{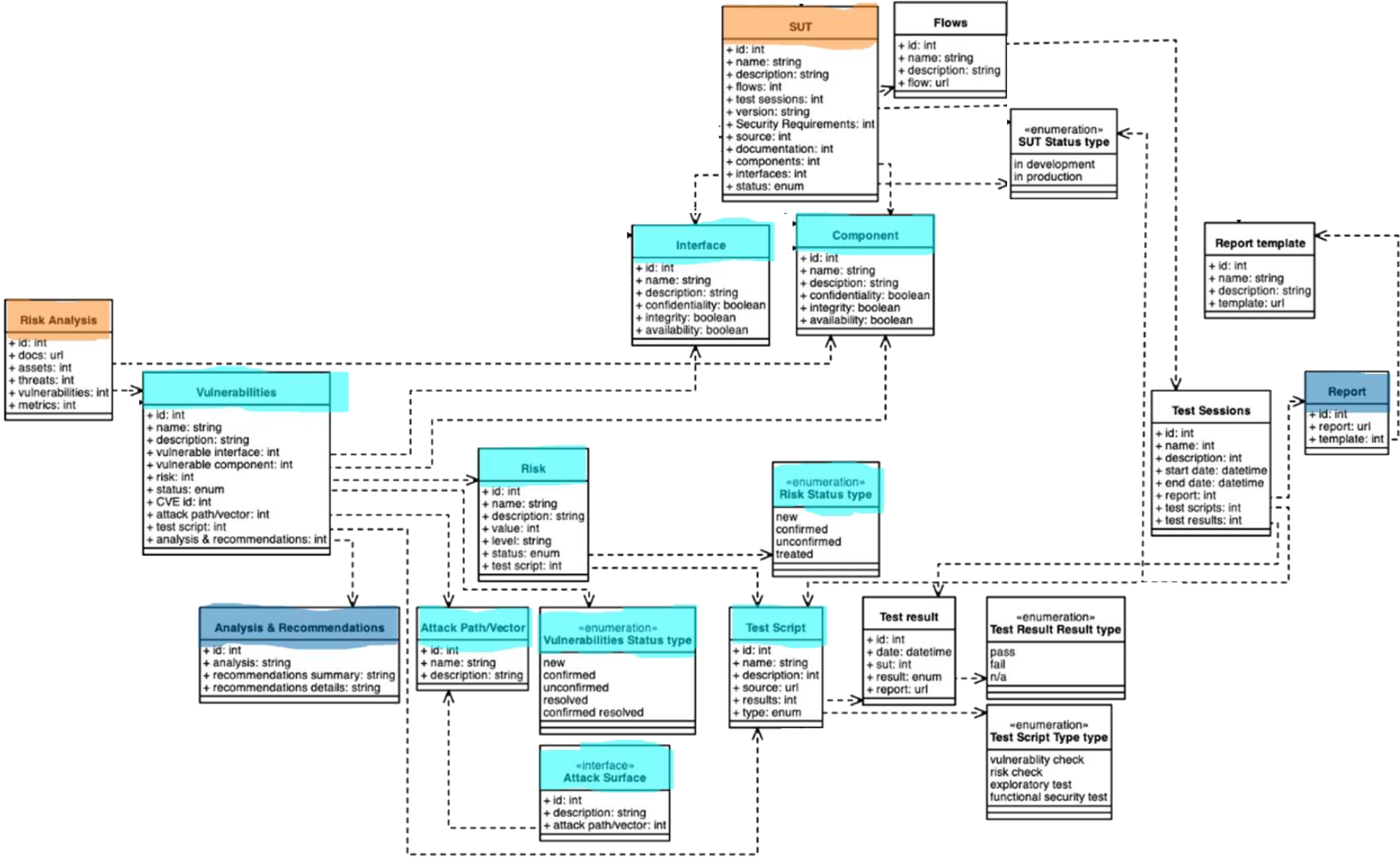}
\caption{CYRUS metamodel (partial)}
\label{fig:MM-CYRUS}
\end{figure*}

\subsection{Capella Tool}

Capella is a system modelling tool implementing the Arcadia methodology \cite{Roques17}. A specific extension was designed to provide a semantic alignment between risk assessment concepts and system modelling concepts \cite{Naouar21}. Figure \ref{fig:MM-naouar} depicts the proposed extension directly implemented as EMF metamodel and that can be queried using an OCL-based query language. Most of the concepts directly match our metamodel. More specific aspects concern the system modelling in terms of functions, interactions and data. A missing notion in our model is the notion of stakeholder which can both act as normal user or threat the system which would maybe require a more precise modelling in Capella but an interesting point is that some agent could have both "cooperative" and "attacker" facets.

\subsection{CYRUS}

Finally, we conducted a test with a R\&D framework supporting test plan generation and reporting \cite{CYRUS}. The tooling is implemented in a relational database with an  web interface and a Birth reporting feature. Figure \ref{fig:MM-CYRUS} depicts a partial view of the CYRUS metamodel. It described the System Under Test (SUT) as a collection of components and interfaces like Capella and thus in a more detailed way as inferred from risk analysis frameworks. Those assets are connected to the risk analysis through their vulnerabilities which leads to attack path and results in test suites which are then traced to test results/reports from test sessions. This makes the connection back to the SUT. Globally, the metamodel is totally sound with our framework which allows us to easily import risks and vulnerabilities from our risk analysis through API calls in MONARC and then inject them in the CYRUS test framework using SQL update statements or higher level API calls.

\section{\uppercase{Discussion}}


The proposed conceptual model offers a reference framework that can first be useful for learning about risk analysis and for transposing similar notions between different frameworks and standards. However, the main contribution concerns the transition to a model-driven risk analysis approach that is more precise and automated than previous manual approaches based on tables and textual documents that quickly show their limits on complex systems.

About the business analysis for estimating impacts, specific modelling notations can implement our metamodel, e.g. we developed a i* extension [REF REMOVED] which captures metamodel concepts related to business assets: goals, agents (zones), dependencies (conduits), attacker, assets (resources), dreaded events (anti-goals) and malicious tasks (attack scenarios). The use of an tool such as piStar immediately enables the  capture these concepts and to make connections with the different standards. The use of higher level model such as the i* strategic model or system model helps to captures the main functional objectives of the system as well as the intention of an attacker. This enable to identify the security properties to be ensured e.g. on data (integrity) or services (availability).

At the technical level, infrastructure diagrams can also be build and mapped to our metamodel especially through the structuring in zones and conduits. The enables automated vulnerability analysis, e.g. based on known attack patterns and CVE (Common Vulnerability Enumeration). Gathering this with the impact information for the business analysis enables an automated generation of the risk matrix and guidance in the identification of countermeasures. It can also support the process of test plan generation and coverage check as shown in our validation. Of course, the more precise the models the more relevant the risks analysis will be but this can also be decided based on the criticality and isolation level of specific sub-systems to also take into account the risk management effort. The integration of the modelling at (infrastructure as) code level also support this, e.g. with tools like Threagile \cite{Threagile}.

Our work share similarities with other modelling effort for risk and security. UML has been extended to capture security properties, for example UMLSec \cite{UMLSec}. The goal-directed requirements engineering methods already mentioned have also been applied more specifically. i* has security extensions for vulnerability analyses \cite{Elahi10}, Secure Tropos allows to deal with socio-technical systems \cite{Paja13} and a security specific ontology has also been proposed \cite{Sales18}. Our metamodel is based on a series of already published metamodels and ontologies reviewed in Section 2.

\section{\uppercase{Conclusion and Perspectives}}

In this paper, we have proposed a metamodel of concepts giving a unified synthesis of the risk analysis approach by reviewing different cyber security frameworks and standards. We have validated it using different tools and show it is globally consistent despite possible improvements in the alignment with some concepts such as the modelling of agents and roles.

Our future work will focus on the refinement of our metamodel as we progress in the implementation of a more elaborated toolchain covering more steps in the DevSecOps lifecycle beyond the risk anakysis and functional testing. We would also like to capture information form penetration testing and from run-time monitoring back into our risk analysis model. Regarding the risk analysis activity, in addition to the conceptual model, we also plan to elaborate a process model for this activity in order to provide a more unified support across variants proposed in different standards.


\bibliographystyle{apalike}
\bibliography{MBRAPT}

\end{document}